\documentclass[useAMS,usenatbib]{mn2e}
\usepackage{graphicx}
\usepackage[usenames]{color}
\usepackage{rotating}
\usepackage{pifont}
\usepackage{amssymb}

\title[The theoretical instability strip of M dwarf stars]{The theoretical instability strip of M dwarf stars}
\author[C. Rodr\'\i guez-L\'opez, J. MacDonald, P. J. Amado, A. Moya, D. Mullan]{C. Rodr\'\i guez-L\'opez$^{1}$\thanks{E-mail:
crl@iaa.es (CRL); jimmacd@udel.edu (JM); pja@iaa.csic.es (PJA); amoya@cab.inta-csic.es (AM);  mullan@bartol.udel.edu (DM)}, J. MacDonald$^{2}$, P.J. Amado$^{1}$, A. Moya$^{3}$ and D. Mullan$^{2}$\\
$^{1}$Dep. de F\'\i sica Estelar. Instituto de Astrof\'\i sica de Andaluc\'\i a (IAA-CSIC), 18008 Granada, Spain \\
$^{2}$Dept. of Physics and Astronomy, University of Delaware, Newark, DE 19716, USA \\
$^{3}$Dep. de Astrof\'\i sica. Centro de Astrobiolog\'ia (CAB, INTA-CSIC), 28691 Villanueva de la Ca\~nada, Madrid, Spain}

\begin{document}

\date{Accepted 2013 December 03. Received 2013 November 28; in original form 2013 October 30}

\pagerange{\pageref{firstpage}--\pageref{lastpage}} \pubyear{2002}

\maketitle

\label{firstpage}

\begin{abstract}
The overstability of the fundamental radial mode in M dwarf models was theoretically predicted by \cite{crl12}. The periods were found to be in the ranges $\sim$25-40~min and $\sim$4-8~h, depending on stellar age and excitation mechanism. We have extended our initial M dwarf model grid in mass, metallicity, and mixing length parameter. We have also considered models with boundary conditions from PHOENIX NextGen atmospheres to test their influence on the pulsation spectra. We find instability of non-radial modes with radial orders up to $k$=3, degree $\ell$=0-3, including p and g modes, with the period range extending from 20~min up to 11~h. Furthermore, we find theoretical evidence of the potential of M dwarfs as solar-like oscillators.
\end{abstract}


\section{Introduction}
 The observational discovery of the first pulsating M dwarf would be a breakthrough in the understanding of the inner structure of the largest population of stars in our Galaxy. The precise measurement of the mass, radius, mean density, size of the convection zone or age, obtained from asteroseismic analysis could also help to address other problems, such as providing constraints for stellar dynamo models \citep{garcia10} and calibrating statistical age-dating methods such as lithium depletion or stellar spin-down.

  So far, observational searches for pulsations in M dwarfs have not been fruitful. The ground base photometric exploration of 120 M dwarfs gathered in \cite{baran11b}, \cite{krzesinski12} and \cite{baran13} attained a precision of about 1~part per thousand with null detections, suggesting that if pulsations are driven and propagate to the surface of the star, their amplitudes may be very low. \cite{baran11a} also explored six M dwarfs observed with the Kepler spacecraft \citep{borucki10} in short cadence (sampling time $\sim$1~min) again with null results, this time with a detection limit of the order of parts per million. We analysed with Fourier techniques 5 Kepler M dwarfs with short cadence data from Kepler Guest Observer program - Cycle 3 ``Pulsations and oversized M dwarfs''  P.I: J. Gizis) attaining detection limits of ppm and finding no significant oscillations. Although not statistically significant given the small sample, these results reinforce the conclusion that if pulsations are to be found, at least precisions of 1~ppm have to be attained.

  \cite{crl12} (RLMM12) provided the first non-radial, non-adiabatic instability study of 0.10-0.50~M$_\odot$ M dwarf models and predicted the excitation of radial modes with periods in ranges of $\sim$25-40~min and $\sim$4-8~h depending on the mass and evolutionary stage. The oscillations were sustained by an epsilon mechanism caused by deuterium (D) or He$^3$ burning for the completely convective models, and by periodic flux blocking at the base of the outer convective envelope for the more massive, partially convective models.

  We present here an extension of the work began by RLMM12 by enlarging the original model grid in mass and mass step, mixing length parameter, metallicity and the use of model atmosphere boundary conditions, to test the influence of these parameters on the frequency spectra. 

  Furthermore, given that M dwarfs are either completely convective or have a large convective envelope, we carry out a theoretical exercise that unveils the potential of M dwarfs to show stochastic oscillations.

  The paper is organised as follows: Section~\ref{sec:models} describes the model grids built for this study; Section~\ref{sec:analysis} deals with the pulsation analysis; Section~\ref{sec:stochastic} discusses the potential of M dwarfs as solar-like oscillators and finally Section~\ref{sec:discussion} presents the discussion and conclusions.

\section{Model grid properties}
\label{sec:models}
 Seven model grids were built with the stellar evolution code described in \cite{macdonald13}, and references therein, to test the influence of the mixing length parameter $\alpha$, the metallicity [Fe/H], and the type of boundary conditions (bcs) on the model pulsation frequencies. The OPAL equation of state \citep{rogers02} was used for all models. For three of the model grids, we have used NextGen atmospheres \citep{hauschildt99, allard00} to determine the outer bcs. Specifically, we use the temperature and pressure at optical depth 100 in the atmosphere models, for given effective temperature ($T_{eff}$), surface gravity (log \textit{g}) and composition. Since these atmosphere models use mixing length ratio, $\alpha = 1$, we have used $\alpha = 1$ in the interior for these models. These grids including NextGen atmospheres differ in metallicity; specifically, we set [Fe/H] = 0, -0.5 and -1. For the four other model grids, the outer bcs are the temperature and pressure at optical depth 0.1 as determined by the Eddington approximation. This set of four grids has solar metallicity but a different value for $\alpha$, specifically from 0.5 to 2.0 in steps of 0.5. For the masses considered here, our models are either fully convective or have deep surface convection zones, which allows us to ignore the effects of element diffusion and gravitational settling. For simplicity, we also ignore the effects of magnetic fields. The model grid characteristics are given in Table~\ref{tab:grids}.

\begin{table}
 \centering
  \caption{Summary of model grids characteristics. The last row gives the colour guide used in most figures.}
  \begin{tabular}{llllllll}
  \hline
   Grid & 1 & 2 & 3 & 4 & 5 & 6 & 7 \\
 \hline
 $\alpha$  & 0.5  & 1.0  & 1.5  & 2.0  & 1.0  & 1.0    & 1.0  \\
 $[Fe/H]$  & 0.0    & 0.0  & 0.0    & 0.0  & 0.0  & -0.5 & -1.0 \\
 ATM       & $X$  & $X$  & $X$  & $X$  & \checkmark  & \checkmark & \checkmark \\
 colour    & red  & green  & blue  & orange  & magenta  & cyan  & purple  \\
\hline
\hline
\end{tabular}
\label{tab:grids}
\end{table}

\begin{figure*}
 \includegraphics[width=130mm,angle=90]{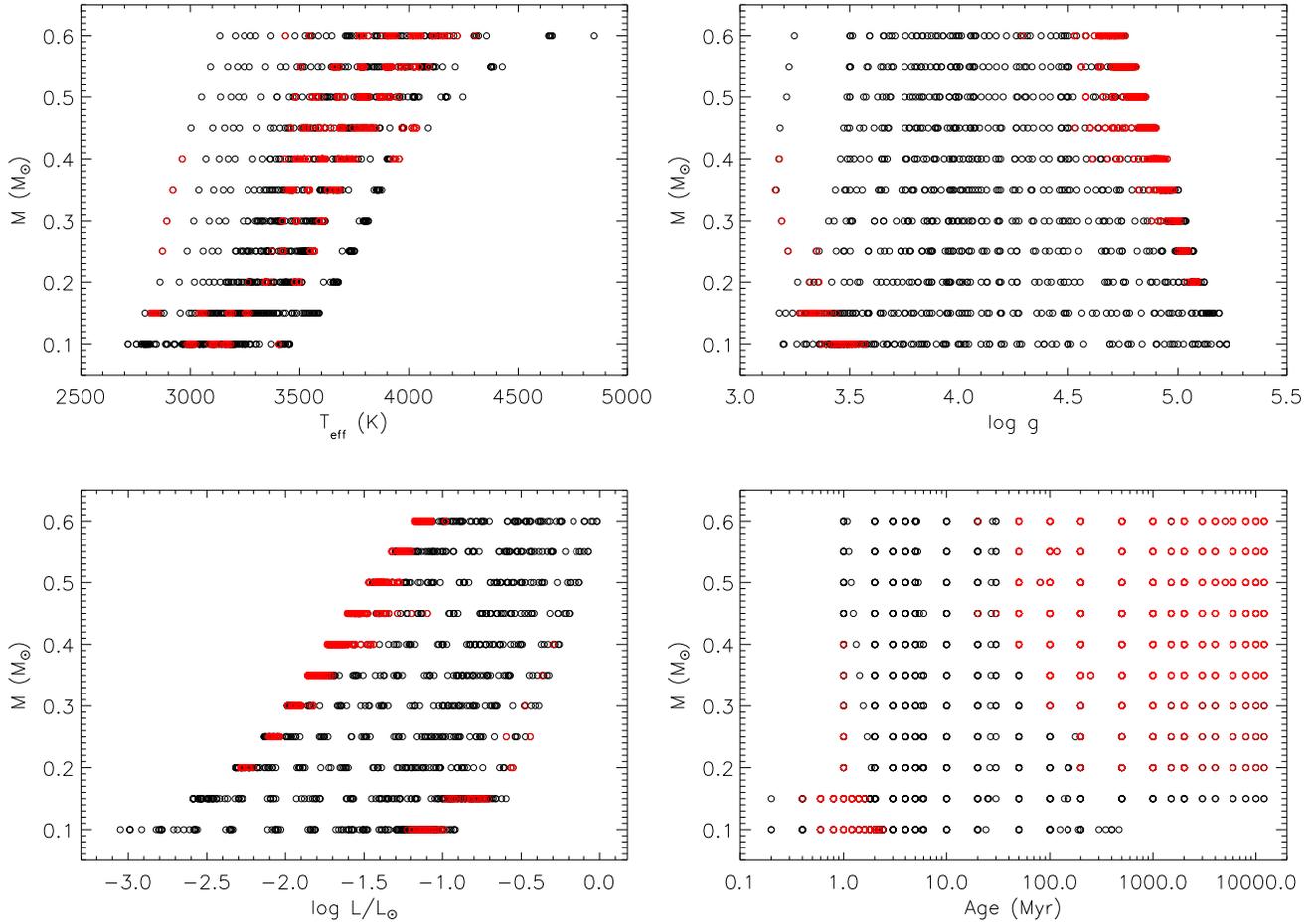}
 \caption{Physical parameters covered by all model grids as a function of the mass. Excited models are shown in red.}
 \label{fig:all_grid_param}
\end{figure*}

 For each model grid, models with masses in the range 0.10 to 0.60~M$_\odot$ in increments of 0.05~M$_\odot$ spanning 12\,000~Myr were calculated. Models are built from the contraction phase and are allowed to evolve up to the age of the Universe, except for the 0.10~M$_\odot$ sequence which terminates when the model structure leaves the OPAL equation of state grid. As H burning in M dwarfs proceeds slowly, the models are practically unevolved once they arrive on the main sequence and remain there for the rest of their evolution. 

  Models from 0.10 to 0.25~M$_\odot$ are totally convective, while the 0.30~M$_\odot$ models already show radiative zones for certain ages. All models on the main sequence with 0.35~M$_\odot$ and higher are partially convective, showing an outer convective envelope and an inner radiative zone. The parameter space in $T_{eff}$, log \textit{g}, luminosity (\textit{L}) and age covered by the whole set of grids and masses is given in Figure~\ref{fig:all_grid_param}. Models in red are found to be unstable.

\begin{figure}
 \includegraphics[width=65mm,angle=90]{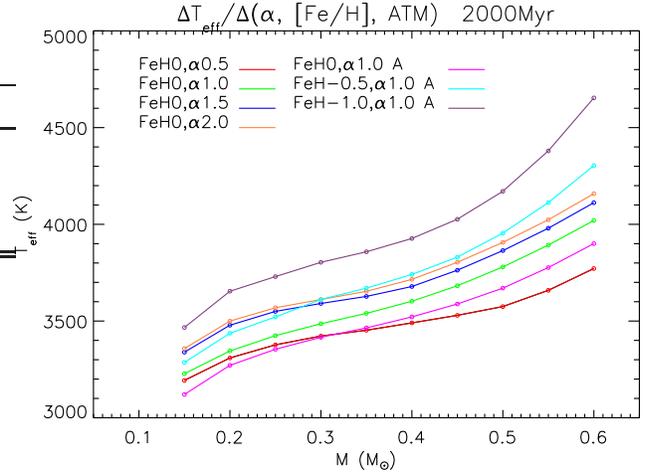}
 \caption{Effective temperature dependence with the $\alpha$ parameter, metallicity and treatment of outer boundary conditions. See the text for more details.}
 \label{fig:teff-mass}
\end{figure}

 Figure~\ref{fig:teff-mass} shows how for a given age, exemplified for 2\,000~Myr models, and mass, increasing the $\alpha$ parameter increases the effective temperature of the model, producing models with lower radii, and as a result, higher log g values. This effect is larger for the higher mass partially convective models, and for values of the ML parameter up to $\alpha$=1.5. As an example, for the 0.60~M$_\odot$ models, $\Delta$T$_{eff}\simeq$400~K between $\alpha$=0.5 and $\alpha$=1.5, but because of saturation of the efficiency of convective energy transfer, $\Delta$T$_{eff}$ is only $\simeq$50~K between $\alpha$=1.5 and $\alpha$=2.0.

From inspection of the atmosphere bcs models, we see an increase in $T_{eff}$  and log g  with decreasing  [Fe/H] at a given $\alpha$, age and mass. The $T_{eff}$ increase is larger for higher masses, going from $\Delta$T$_{eff}\simeq$350~K for the 0.15~M$_\odot$ models to $\Delta$T$_{eff}\simeq$750~K for the 0.60~M$_\odot$ models with [Fe/H]=0.0 and -1.0. This behaviour is a consequence of the lower mass models having denser atmospheres than the higher mass models, which leads to the convective efficiency being closer to saturation in the lower mass models, for the same value of $\alpha$ and [Fe/H]. In addition, reducing [Fe/H] also leads to denser atmospheres and greater convective efficiency. Thus the differential effect of reducing [Fe/H] on $T_{eff}$ is larger for the higher mass stars.

\begin{figure}
 \includegraphics[width=65mm,angle=90]{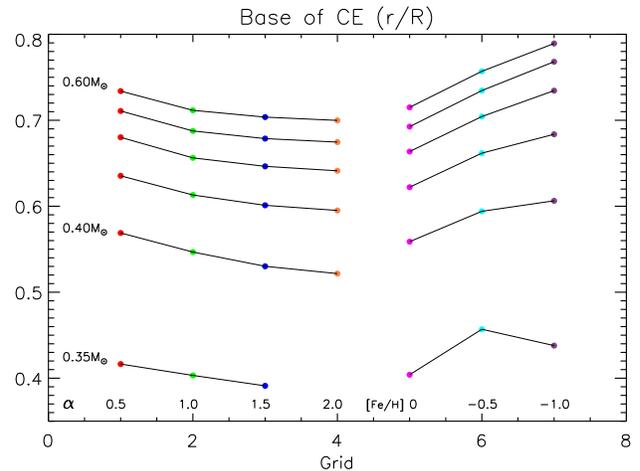}
 \caption{Grid dependence of the base of the convective envelope for 2\,000Myr partially convective models. See the text for details.}
 \label{fig:ce_age2000}
\end{figure}

 We also see that the atmosphere bcs models have lower $T_{eff}$  and log g than the Eddington bcs models with the same [Fe/H] and ML parameter ([Fe/H]=0 and $\alpha$=1 for grids 2 and 5 -green and magenta tracks-). The decrease in $T_{eff}$ is about $\Delta$T$_{eff}\simeq$100~K for the whole mass range.

 Figure~\ref{fig:ce_age2000} shows the mass dependence of the position of the base of the convective envelope (CE) for 2\,000Myr partially convective models. At this age the stars are slowly evolving on the Main Sequence and the results shown in Fig.~\ref{fig:ce_age2000} can be considered representative for all of the Main Sequence. Increasing $\alpha$ yields larger, i.e. deeper, convective envelopes, while  decreasing the metallicity  or use of atmosphere bcs (magenta grid compared to the green one) yields shallower CEs.

\section{Pulsation Analysis}
\label{sec:analysis}
 The pulsational instabilities were calculated with the GRACO non-radial, non-adiabatic pulsation code \citep{moya08,moya04}. We explored modes with degree $\ell$=0 to $\ell$=3 and frequencies from 20~$\mu$Hz to the acoustic cut-off frequency of the model, which may be as high as 40\,000~$\mu$Hz. The code has the possibility of including the convection-pulsation interaction through the time dependent convection (TDC) formulation \citep{dupret04}. However, as pulsation periods were found to be much shorter than the local convective time scale, the TDC implementation is not needed and the frozen convection approximation that we use is justified.

 The work integral ($W$) and work derivative ($dW$) evaluated throughout the star give the contribution of every layer to the energy balance of a mode. Careful analysis of these quantities shows that, for some modes, there is a large contribution from the very outer layers, less than 0.1\% in fractional radius, that outnumbers any other driving or damping region within the star. When this happens the growth rate, $\eta= W/\int_0^R|dW/dr|dr$ \citep{stellingwerf78}, is highly dependent on the choice of the outer bcs, as it is dominated by the numerical noise of the outer layers. 

  To get rid of artificially excited modes, the models were subjected to the pulsation analysis with two different bcs: the outer mechanical condition described in \cite{unno89} for a finite pressure at the outermost layer and $\delta P=0$. If the model is to be excited, the growth rate should be independent of the chosen pulsational bcs. As will be seen below, it was found that the grid with $\alpha$=0.5 and the atmosphere bcs grids were especially sensitive to the choice of pulsational bcs. This is due to the higher density of the very outer layers, which then have a higher weight in the energy balance of the modes. This dependence of the $\alpha$=0.5 and the atmosphere bcs grids on the bcs gives rise to all modes being artificially excited for certain masses.

  In the following sections we analyse the results of the pulsation analysis for the mass ranges with different convection zone behaviours.

\subsection{Fully convective models: 0.10-0.25~M$_\odot$ models}
  A summary of the outcome of the pulsation analysis for the fully convective models is given in Table~\ref{tab:0.10-0.25}. Each column shows for the different grids, the periods, \textit{e}-folding times, ages, excitation mechanism and the degree and radial order of the excited modes.

  Only the fundamental mode ($\ell$=0, $k$=0), which has large amplitude in the core and overcomes the damping of external regions, is excited due to the driving produced by the $\epsilon$ mechanism associated with D or He$^3$ burning.

  The \textit{e}-folding time, $\tau_{fold}$, defined as the inverse of the absolute imaginary part of the eigenfrequency, gives the time needed to increase the initial amplitude of the mode by a factor of \textit{e}. As the amplitude of the mode is an unknown in linear oscillation codes, such as the one used here, the \textit{e}-folding time yields an estimate of the observability of the mode. Accordingly, the shorter the $\tau_{fold}$ compared with the age of the excited model, the higher the probability of the mode to be detected, as it allows more time for the amplitude to grow.

  The models excited by D burning are very young, still in the contraction phase, with periods in the 4 to 9~h range. Their $\tau_{fold}$ values are of the order of or larger than the age of the models, meaning that modes are not favoured for detection unless their initial (unknown) amplitudes are large. The models excited by He$^3$ burning are already on the main sequence and show much shorter periods, in the 20-30~min range, and have much more favourable $\tau_{fold}$ values for the modes to develop observable amplitudes.

  As explained above, the $\alpha$=0.5 and atmosphere bcs grids not listed were subject to large numerical noise in the outer layers that only produced artificially excited modes.

\subsection{Transition to partially convective models: 0.30~M$_\odot$ models}
 The summary of the pulsation analysis of what we call transition models is given in Table~\ref{tab:0.30}. 

  The 50-100~Myr 0.30~M$_\odot$ models have an inner radiative zone and due to the large convective time scale at the tachocline, the convective flux cannot promptly adjust to transport the radiative flux coming from below. This periodic flux blocking at the base of the convective envelope effectively drives the oscillations of the 100~Myr models, just about to enter the main sequence, and justifies the frozen convection approximation used in the oscillation code.

 Models still show instabilities caused by the D burning at age 1~Myr, which yields periods of 10~h, and by He$^{3}$ burning on the main sequence. Modes unstable due to He$^3$ burning and flux blocking mechanisms have periods in the 30-35~min range. Again, as for the lower mass models, only the fundamental radial mode is excited and driving based on D or flux blocking may not be efficient due to the long e-folding times compared to the age of the models.

\begin{figure}
 \includegraphics[width=65mm,angle=90]{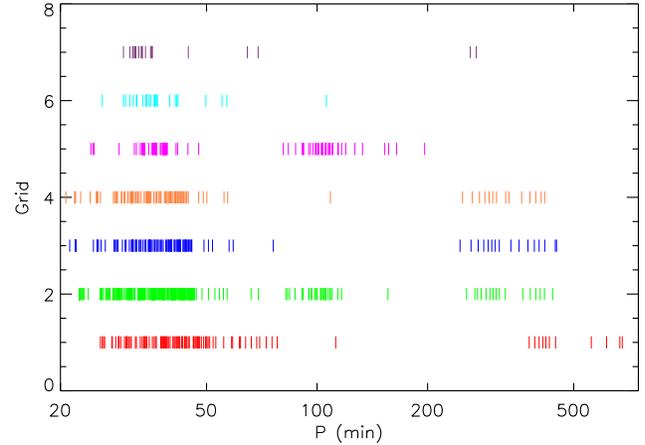}
 \caption{Grid dependence for the excited periods found in the full range of masses. See the text for details.}
 \label{fig:all_periods}
\end{figure}

\subsection{Partially convective models: 0.35-0.60~M$_\odot$ models}
  The summary of the pulsation analysis for the partially convective models is given in Table~\ref{tab:0.35-0.60}. We point out that the e-folding times given are for the fundamental radial mode, which are longer than those of the $\ell$=2, $k$=0 modes, especially for the younger models, where it is more restrictive, but we prefer to be conservative.

  The flux blocking mechanism, described in the previous section, excites models as young as 20~Myr and all along the evolutionary track. The fundamental ($k$=0) radial and non-radial modes ($\ell$=0 - 3) are excited as well as the $k$=1,2 ($\ell$=1 - 3) pressure modes, labeled p1,p2 in the table. Periods lie mostly in the 20-60~min range, except for the younger models, less likely to be observed due to $\tau_{fold}$ values comparable to the age of the model, in which periods cover the 1-2~h range.

  Gravity modes with $\ell$=1, 2 and $k$=-1, -2, -3, labeled g1, g2, g3 in the table, are excited with periods in the 1-3~h range. Driving is due to a combination of the He$^3$ burning $\epsilon$ mechanism and the flux blocking mechanism, with the latter usually prevailing to excite the mode just by itself.

  The D burning $\epsilon$ mechanism is responsible for driving only for the $\alpha$=0.5, 1~Myr models, yielding excited periods of 11~h, although with a low probability of being excited due to long e-folding times.

  The failure to excite modes of the 0.50 to 0.60~M$_\odot$ atmosphere bcs models is due to their sensitivity to the outer pulsational bcs.

\subsection{The instability strip}

  Figure~\ref{fig:all_periods} shows the whole range of excited periods for each grid. Although grid dependent, the 20-500 minute range appears to be covered by some excited model. Table~\ref{tab:summary} is a summary of the pulsation analysis sorted by the pulsation period, intended  to be useful for prospective observers in search of M dwarf pulsations. It segregates the excited models by pulsation period and gives details about the corresponding mass ranges and excitation mechanism.

\begin{figure}
 \includegraphics[width=65mm,angle=90]{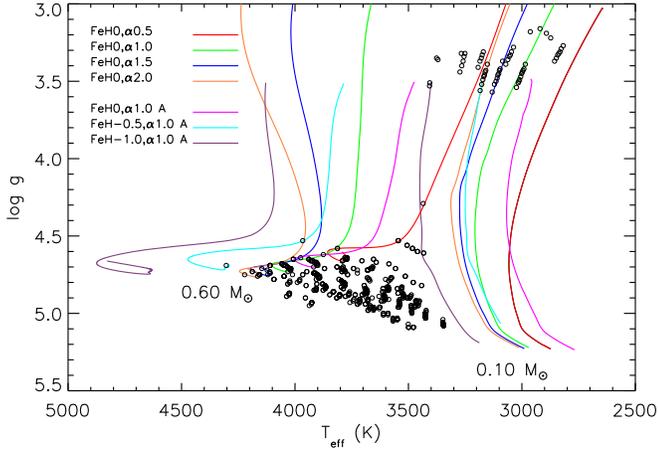}
 \caption{Evolutionary tracks of the 0.10 and 0.60M$_\odot$ models of each grid. Open circles plot the unstable models and define the instability strip in terms of T$_{eff}$ and log g.}
 \label{fig:tracks}
\end{figure}

  Figure~\ref{fig:tracks} shows the evolutionary tracks of bracketing 0.10 and 0.60M$_\odot$ models corresponding to each grid. All other evolutionary tracks for 0.15 to 0.55M$_\odot$ are contained within the plotted ones and are not shown for clarity. The excited models are plotted with black open circles and delimit two islands of instability within the following ranges: log \textit{g} = [3.1,3.6], $T_{eff}$=[2800,3500] whose boundaries are set by the short lifetimes of D burning; and log \textit{g} =[4.5,5.1], $T_{eff}$=[3300,4300] for the models excited by the He$^{3}$ burning $\epsilon$ and flux blocking mechanisms. The onset of the instability is produced once the flux-blocking mechanism is strong enough to overcome the damping of the inner layers of the star.

   Figure~\ref{fig:p_fits} shows the dependence of the period of the fundamental radial mode ($\ell$=0, $k$=0) for the 0.20 to 0.60M$_\odot$ of 2\,000~Myr models with the fundamental parameters: $T_{eff}$, log $L/L\odot$, log \textit{g}, radius, size of the convective envelope and mean density of the models. This is shown for all grids (each colour corresponds to a different grid, see Table~\ref{tab:grids}), i.e. we can see the effect of changing $\alpha$, metallicity and outer bcs on the periods:

\begin{figure*}
 \includegraphics[width=130mm,angle=90]{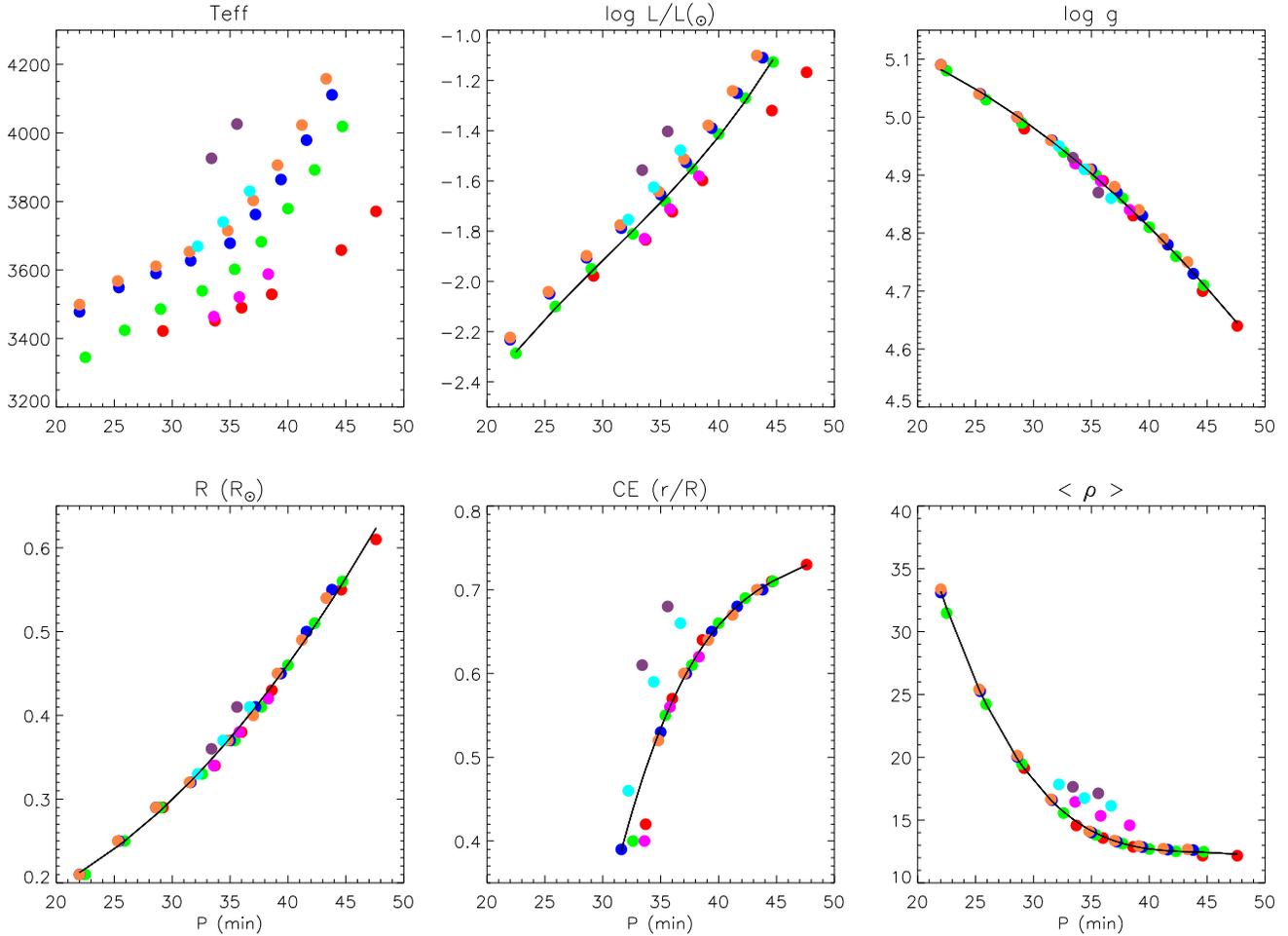}
 \caption{Period dependence of the fundamental radial mode with physical parameters for the 0.20 to 0.60M$_\odot$ models of 2\,000~Myr for all grids. See text for details.}
 \label{fig:p_fits}
\end{figure*}

\begin{table}
 \setlength\tabcolsep{4pt} 
 \centering
  \caption{Summary of pulsation analysis sorted by pulsation period. The last column gives information on the likelihood of the observability of a mode: when the e-folding time is shorter than the age of the model, the observational detection of pulsations is more likely. This is true for models older than 50~Myr. See the text for details.}
  \begin{tabular}{lllll}
  \hline
   Mass         & P     & Excit.  & Comments &  e-fold $<$  \\
   (M$_\odot$)  &       & mechan. &          &        age        \\
 \hline
 0.10 - 0.40  & 4 - 11~h      & $\epsilon$ (D)           &  age $\lesssim$ 2~Myr &  $X$ \\
              &               &                          &                       &  \\
 0.40 - 0.60  & 1 - 2~h       & F-b                      & age $\leq$ 50~Myr  &  $X$  \\
              & 1 - 3~h       & F-b, $\epsilon$ (He$^3$) & g-modes & \checkmark \\
              &               &                          &         &  \\
 0.20 - 0.30  & 20 - 30~min   & $\epsilon$ (He$^3$)      &         &   \checkmark   \\
 0.35 - 0.60  & 20 - 60~min   & F-b                      &         &   \checkmark \\
\hline
\hline
\end{tabular}
\label{tab:summary}
\end{table}

\begin{itemize}
 \item the ``vertical`` colour sequences (each coloured dot corresponds to a different grid) in the upper left plot are mass sequences, starting with the 0.20M$_\odot$ models with about 23~min periods and going up to the 0.60M$_\odot$ models within the 43-49~min range period. Thus knowledge of the period could provide an independent measure of the total mass of the model. We recall here that grids varying with $\alpha$ from 0.5-2.0 correspond to red, green, blue and orange dots, so, as we saw in Figure~\ref{fig:teff-mass}, increasing $\alpha$ increases the $T_{eff}$ of the model and a decrease in period is produced. Decreasing [Fe/H] (magenta, cyan and violet dots) works in the same way, while use of atmosphere bcs (magenta to green model) has the opposite effect.

  Having periods measured with a precision of a second, and also having an accurate spectroscopic determination of the $T_{eff}$ may enable observers to choose between the different grids, advancing our knowledge of the ad-hoc $\alpha$ parameter and yielding an independent determination of [Fe/H].

  \item from the upper middle plot we see that there is a grid dependent period-luminosity relation. As an example, we have fitted 1 to 3 degree polynomials to the solar metallicity, $\alpha$=1 grid results (green dots). The black line shows the best cubic fit: 

  \[\log L/L_{\odot} = -4.62517 + 0.177995P - 0.00434748P^2\]
   \[+ 4.74567~10^{-5}P^3\]

where $P$ stands for the period in minutes.

 The period increases with increasing luminosity and an independent determination of the mass can again be obtained from the period.

  \item the line in the upper right panel is the best quadratic fit to log \textit{g}: 

  \[\log g = 5.18522 + 0.00102656P - 0.000260110P^2\]

  This is grid independent, so knowledge of the period automatically gives an independent measure of log \textit{g} that can be further compared to spectroscopic determinations. We again note that increasing $\alpha$ increases log \textit{g}.

  \item the line in the lower left panel shows the best quadratic fit to the radius:

  \[R/R_{\odot} = 0.168510 - 0.00449263P + 0.000295172P^2\]

  so, again the period gives an independent and direct measure of the radius. This may be extremely useful in unravelling the long standing problem of the discrepancies between observed and theoretically predicted M dwarf radii.

  \item the dependence of the location of the base of the convective envelope can be also tested by the period (lower middle panel), which is grid independent for the solar [Fe/H] models. Shorter periods correspond to deeper envelopes and lower masses. The best fit is given by the cubic polynomial:

  \[\mathrm{CE~base}~(r/R) = -6.06713 + 0.417345P - 0.00866586P^2\]
   \[+ 6.08777~10^{-5}P^3\]

  Decreasing [Fe/H] clearly produces shallower envelopes and shorter periods. 

  \item finally the lower right panel shows how the period gives an independent measure of the mean density (cgs units) of the model, which is grid independent for the solar metallicity models and is given by the cubic fit:

\[<\rho> = 171.970 - 10.7567P + 0.242380P^2 - 0.00182505P^3\]

  Periods of the fundamental radial mode decrease with increasing density, following the period-mean density relation, $P\approx \sqrt{3\pi /2 \gamma G \rho}$. 

  Lower metallicities give denser models and shorter periods.

\end{itemize}

In general, the period increases with increasing mass and, for a given mass, the period increases with increasing metallicity, decreasing $\alpha$ and for models with an atmosphere.

Models with masses larger than 0.30M$_\odot$ can excite not only the fundamental radial mode, but also other modes, being the most frequent the $\ell$=2 and $\ell$=3, $k$=0 modes, as shown in Table~\ref{tab:0.35-0.60}. For these two modes, the plots in Figure~\ref{fig:p_fits} are replicated, but the periods would be shifted with respect to the $\ell$=0, $k$=0 mode by +5~minutes for the $\ell$=2, $k$=0 and by -5~minutes for the $\ell$=3, $k$=0 mode. For the sake of clarity, they are not shown in Figure~\ref{fig:p_fits}. If pulsations are observationally found, the mode with largest amplitude is usually the fundamental radial mode, but in the case of multimode pulsators, a modal identification is due, although out of the scope of this paper.

\section{Could stochastic modes be driven in M dwarf stars?}
\label{sec:stochastic}
  While our non-adiabatic oscillations code evaluates only thermodynamical effects to assess the overstability of a mode, we still can gain some insight into the theoretical feasibility of stochastically excited modes. These modes are driven by the turbulent motions produced in the convection zones of stars like our Sun. Given that M dwarfs have an outer convection zone or even are completely convective, they have the potential to drive these modes.

  Following the discussion by \cite{mullan10} on the selection mechanism for excited solar modes, namely that a pressure mode can be driven more effectively if its largest antinode is located within the convection zone and in the region of maximum available energy to be invested in sound waves, we computed the flux of the energy in sound waves, $F_s$, of each model as:

\begin{equation}
 F_{\rmn{s}} \sim F_{\rmn{p}} \left({{V} \over {c_{\rmn{s}}}}\right)^5 \sim \rho V^3 M^5
\end{equation}

  where $F_p$ is the maximum rate at which the kinetic energy per unit volume of granule flow can be released into the surrounding medium from a granule which survives for a single convective turnover time, $V$ is the convective velocity, $c_s$ is the sound speed and $M$ is the Mach number.

  We consider that a mode has the potential to be stochastically excited if:

  1/ The last antinode of the $r \rho^{1/2} \delta r$ function, which gives the contribution at a given radius to the energy density, is the largest and lies in the convection zone (which is always fulfilled for any of our models) and

  2/ The location of this last antinode falls within the full width of half maximum of the acoustic flux. This is the more restrictive condition. We could relax it, if we wanted more modes to survive this cut.

  In Table~\ref{tab:stochastic2}, we summarize the period range of possible stochastically excited modes for solar metallicity models and values of the mixing length parameter $\alpha=$2. Modes where calculated from $\ell=$0 to $\ell=$2 up to the cut-off frequency of each model.

\begin{figure}
 \includegraphics[width=65mm,angle=90]{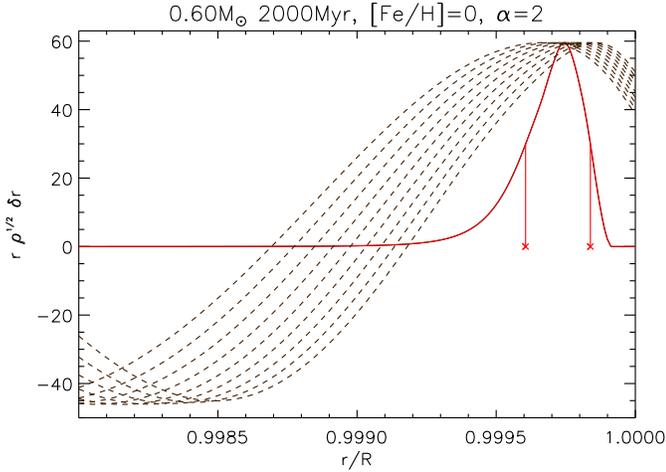}
 \caption{The acoustic flux ($F_s$) is plotted with a red solid line. The vertical red lines mark its full width at half maximum. If the largest antinode of the $r \rho^{1/2} \delta r$ function falls within this region, the mode has the potential to be stochastically excited. For this particular model, modes in the range $k$=[29,38] are selected and plotted with dashed lines. The contribution to the energy density $r \rho^{1/2} \delta r$ is normalized and then scaled to the $F_s$ maximum. Note that only the 0.2\% of the outer fractional radius is plotted.}
 \label{fig:stochastic}
\end{figure}

\begin{table}
\caption{Radial orders ($k$) and period ranges of potentially stochastically excited modes for solar metallicity models and mixing length parameter $\alpha$=2. The value of the $F_s$ maximum is also given. See the text for more details.}
\label{tab:stochastic2}
\centering
\renewcommand{\footnoterule}{}
\renewcommand{\tabcolsep}{5pt}
\begin{tabular}{ccccc}
\hline
   M          & Age    &  $P$    & $k$ &  max(F$_s$)  \\
  (M$_\odot$)  & (Myr)  & (min)   &       &  erg/cm$^2$ s\\
\hline
 0.10  & 0.4 - 20  & 40 - 2  & [12,15], [26,28]  & 20-2   \\
       & 50 - 300  &  $<$2   & [31,34], [41,46]  & 1-0.1  \\
 \\
 0.15  & 0.4 - 20  & 58 - 2  & [9,15], [28,30]   & 20-5   \\
       & 50 - 12000&   $<$2  & [34,35], [43,45]  & 3-1    \\
 \\
 0.20  & 1 - 20    & 48 - 2  & [10,14], [28,31]  & 50-5   \\
       & 50 - 2000 &  $<$2   & [34,36], [43,45]  & 4-2    \\
 \\
 0.25  & 1 - 20    & 40 - 2  & [11,14], [28,32]  & 90-5   \\
       & 50 - 2000 &  $<$2   & [34,37], [41,45]  & 4-2    \\
 \\
 0.30  & 1 - 20    & 34 - 2  & [12,15], [28,33]  & 145-6  \\
       & 50 - 2000 &  $<$2   & [34,38], [42,45]  & 4-3    \\
 \\
 0.35  & 1 - 20    & 31 - 2  & [13,16], [28,33]  & 220-7  \\
       & 50 - 2000 &  $<$2   & [34,39], [40,44]  & 4-3    \\
 \\
 0.40  & 1 - 20    & 31 - 2  & [13,17], [25,34]  & 300-8  \\
       & 50 - 2000 &  $<$2   & [33,39], [39,44]  & 5-3    \\
 \\
 0.45  & 1 - 20    & 29 - 2  & [13,18], [25,34]  & 365-10 \\
       & 50 - 2000 &  $<$2   & [33,39], [38,43]  & 6-4    \\
 \\
 0.50  & 1 - 20    & 29 - 2  & [13,18], [25,34]  & 430-15 \\
       & 50 - 2000 &  $<$2   & [30,39], [33,42]  & 8-7    \\
 \\
 0.55  & 1 - 20    & 29 - 2   & [14,18], [25,35]  & 585-20 \\
       & 50 - 2000 &  $<$2    & [28,38], [30,40]  & 15-18  \\
 \\
 0.60  & 1 - 50    & 28 - 2   & [14,17], [28,37]  & 770-40  \\
       & 100 - 2000&  $<$2    & [28,37], [29,38]  & 85-60   \\
\hline
\end{tabular}
\end{table}

  We group models with periods larger than 2~min, with the potential of being observationally detectable with short cadence Kepler spacecraft \citep{borucki10} data, and models with periods below 2~min. The bracketed $k$ values correspond to the radial order of excited modes for each model listed under the 'Age' column. The possible stochastically excited modes with periods $>$2~min correspond to very young models ($\leq$ 50~Myr).

  The location and width of the acoustic flux maximum, $F_s$, select radial orders in the range from about $k=$10 for the youngest models to about $k=$40 for the oldest ones. For each individual model, only about 2 to 10 radial orders are selected. In the table, the maximum value attained by F$_s$ is also given, which gives a flavour of the likelihood of excitation of a mode. 
 
  In the case of a solar model, \cite{mullan10} showed that differences of up to three orders of magnitude in the F$_s$ function at the location of the solar pressure modes with $k=$25, 15 and 10 could account for the differences in power observed for those modes. As a comparison, in our study, for the 0.60~M$_\odot$, $\alpha=$2 models, the maximum values of the $F_s$ function go from 770 to 60~erg\,cm$^{-2}$\,s$^{-1}$ for the 1~Myr and 2\,000~Myr model, respectively; so although the location and width of $F_s$ certainly has a selection effect of modes, selecting 29$ \leq k_r\leq$38 for the 2\,000~Myr model, its magnitude may not give much room to further culled modes within these high radial orders. This is illustrated in Figure~\ref{fig:stochastic} which plots, for that model, the scaled eigenradial displacement function quantity $r \rho^{1/2} \delta r$ in the region of maximum F$_s$, which for all analysed models lies at a fractional radius less than 0.2\% from the surface. For clarity, only the $\ell =2$ modes are plotted, although similar plots were obtained for other $\ell$'s.

  We have done the same analysis for the solar metallicity $\alpha=$1 model grid. In general, periods are somewhat longer, $F_s$ values are lower and similar $k$ are selected. 

  Regarding the observational detectability of the stochastic modes, to be able to detect those with periods $\simeq$2~min, exposure times of a maximum of half the period, i.e. $\sim$1~min, but preferably shorter, are needed. Kepler spacecraft short cadence data have exposure times of 58.85~s which place the expected oscillation period around the Nyquist frequency, where the short cadence data have at least four spurious frequencies. All this, added to the intrinsic faintness of the objects and the expected low amplitude oscillations, makes the observational detection a real, but 'should-be-taken', challenge.

  Models with periods $>$2~min correspond to very young M dwarfs ($\leq$ 50~Myr) still in their pre-Main Sequence contraction phase. They may lie in stellar formation regions and be embedded in dust and gas, which may render them undetectable to optical studies from the ground. These pulsation searches could make an optimal case for the James Webb Space Telescope\footnote{http://webbtelescope.org/webb\_telescope/} which is being designed to operate mainly in the infrared and having as one of its scientific cornerstones peering into star forming regions to study star and planet formation or to the high resolution spectrograph HIRES on the E-ELT~\citep{maiolino13}. 

  If observationally detected, there would be no confusion between stochastically or thermodynamically excited modes: in the 20-40~min range stochastically excited modes correspond to very young models, still in the contracting phase, while young models thermodynamically excited by the $\epsilon$ mechanism have much longer periods. If models are on the main sequence, stochastic oscillations have periods shorter than two minutes, impossible to be mistaken for thermodynamical excitation.

\section{Discussion and conclusions}
\label{sec:discussion}
 We present a pulsation analysis of M dwarf models in the 0.10 to 0.60~M$_\odot$ range with different parameters of mixing length, metallicity and treatment of outer boundary condition. We find excitation of the fundamental radial mode as well as non radial p- and g-modes close to the fundamental one. The excitation mechanisms at work to drive the oscillations are the $\epsilon$ mechanism linked to nuclear burning of D or He$^{3}$ and convective flux blocking at the tachocline for partially convective models.

  We give a useful guide for observers to the sampling times needed to detect the pulsations or, in case of the exoplanetists, to be avoided to conceal the pulsation signatures in the light curves to search for planets. Unstable modes can be found in the period range from 20~min to 11~h depending on the grid, mass and evolutionary stage of the model.

  We also show how from the pulsation period an independent measure of spectroscopic quantities such as effective temperature, log \textit{g} and luminosity can be obtained; but most important independent measures of mass, radius, mean density and size of the convective envelope. Moreover, the precise knowledge of the pulsation period, together with an accurate spectroscopic determination of the effective temperature may be enough to constrain the ad-hoc mixing length parameter $\alpha$, metallicity and outer boundary conditions.

  Only with the observational detection of pulsations can an instability strip be further refined, by building a dense grid of models in the vicinity of the spectroscopic box of physical parameters determined for a particular object of interest. As the effect of the different mixing length parameter, metallicity or even the use of atmosphere boundary conditions in frequency spectra is subtle, only by comparing theory and observations can we get reliable answers to the real values of these parameters.

  Our team pursues the detection of theoretically predicted pulsations in M dwarfs from fast time-series spectroscopy using the high-resolution echelle spectrograph HARPS\footnote{http://www.eso.org/sci/facilities/lasilla/instruments/harps.html}. The high precision attainable, under 1~m/s using data analysis software like HARPS-TERRA \citep{anglada12} would enable the detection of low amplitude oscillations in M dwarfs if they are observable.

  Another optimum instrument to search for pulsations around M dwarfs is the high resolution optical and near infrared spectrograph CARMENES \citep{quirrenbach12} whose expected first light is 2015. It is primarily designed to search for Exo-Earths around M dwarfs, but one of the secondary scientific cases is the search for oscillations around M dwarfs.

  Finally, we give for the first time theoretical evidence that solar-like oscillations can be excited in M dwarf models. The observational detection of these modes presents itself as a challenge either because the excited models are predicted to be very young, possibly in an accretion phase obscured by dust and gas, or due to the extremely fast oscillations below two minutes. We may need very precise technology to be able to detect these oscillations, but who said that Nature would reveal her mysteries so easily? Science is only for the bold!

\section*{Acknowledgments}
CR-L has a post-doctoral contract of the JAE-Doc program ``Junta para la ampliaci\'on de Estudios`` (CSIC) co-funded by the FSE (European Social Fund) and acknowledges financial support from the Annie Jump Cannon Fund of the Department of Physics and Astronomy, University of Delaware. CR-L and PJA acknowledge financial support from grants AYA2011-14119-E and AYA2011-30147-C03-01 of the Spanish Ministry of Economy and Competivity (MINECO), co-funded with FEDER funds, and 2011 FQM 7363 of the Consejer\'\i a de Econom\'\i a, Innovaci\'on, Ciencia y Empleo (Junta de Andaluc\'\i a). JM acknowledges financial support from NASA Kepler GO Cycle 3 grant NNX12AAC90G.



\bibliographystyle{aa}
\bibliography{biblio}

\begin{thebibliography}{20}
\expandafter\ifx\csname natexlab\endcsname\relax\def\natexlab#1{#1}\fi

\bibitem[{{Allard} {et~al.}(2000){Allard}, {Hauschildt}, \&
  {Schweitzer}}]{allard00}
{Allard}, F., {Hauschildt}, P.~H., \& {Schweitzer}, A. 2000, \apj, 539, 366

\bibitem[{{Anglada-Escud{\'e}} \& {Butler}(2012)}]{anglada12}
{Anglada-Escud{\'e}}, G. \& {Butler}, R.~P. 2012, \apjs, 200, 15

\bibitem[{{Baran} {et~al.}(2011{\natexlab{a}}){Baran}, {Fox-Machado}, {Lykke},
  {Nielsen}, \& {Telting}}]{baran11a}
{Baran}, A.~S., {Fox-Machado}, L., {Lykke}, J., {Nielsen}, M., \& {Telting},
  J.~H. 2011{\natexlab{a}}, \actaa, 61, 325

\bibitem[{{Baran} {et~al.}(2011{\natexlab{b}}){Baran}, {Winiarski},
  {Krzesi{\'n}ski}, {Fox-Machado}, {Kawaler}, {Dr{\'o}{\.z}dz}, {Faltenbacher},
  {Thompson}, \& {Reed}}]{baran11b}
{Baran}, A.~S., {Winiarski}, M., {Krzesi{\'n}ski}, J., {et~al.}
  2011{\natexlab{b}}, \actaa, 61, 37

\bibitem[{{Baran} {et~al.}(2013){Baran}, {Winiarski}, {Siwak}, {Fox-Machado},
  {Kozie{\l}-Wierzbowska}, {Krzesinski}, {Dr{\'o}{\.z}dz}, \&
  {Winans}}]{baran13}
{Baran}, A.~S., {Winiarski}, M., {Siwak}, M., {et~al.} 2013, \actaa, 63, 41

\bibitem[{{Borucki} {et~al.}(2010){Borucki}, {Koch}, {Basri}, {Batalha},
  {Brown}, {Caldwell}, {Caldwell}, {Christensen-Dalsgaard}, {Cochran},
  {DeVore}, {Dunham}, {Dupree}, {Gautier}, {Geary}, {Gilliland}, {Gould},
  {Howell}, {Jenkins}, {Kondo}, {Latham}, {Marcy}, {Meibom}, {Kjeldsen},
  {Lissauer}, {Monet}, {Morrison}, {Sasselov}, {Tarter}, {Boss}, {Brownlee},
  {Owen}, {Buzasi}, {Charbonneau}, {Doyle}, {Fortney}, {Ford}, {Holman},
  {Seager}, {Steffen}, {Welsh}, {Rowe}, {Anderson}, {Buchhave}, {Ciardi},
  {Walkowicz}, {Sherry}, {Horch}, {Isaacson}, {Everett}, {Fischer}, {Torres},
  {Johnson}, {Endl}, {MacQueen}, {Bryson}, {Dotson}, {Haas}, {Kolodziejczak},
  {Van Cleve}, {Chandrasekaran}, {Twicken}, {Quintana}, {Clarke}, {Allen},
  {Li}, {Wu}, {Tenenbaum}, {Verner}, {Bruhweiler}, {Barnes}, \&
  {Prsa}}]{borucki10}
{Borucki}, W.~J., {Koch}, D., {Basri}, G., {et~al.} 2010, Science, 327, 977

\bibitem[{{Dupret} {et~al.}(2004){Dupret}, {Grigahc{\`e}ne}, {Garrido},
  {Gabriel}, \& {Scuflaire}}]{dupret04}
{Dupret}, M.-A., {Grigahc{\`e}ne}, A., {Garrido}, R., {Gabriel}, M., \&
  {Scuflaire}, R. 2004, \aap, 414, L17

\bibitem[{{Garc{\'{\i}}a} {et~al.}(2010){Garc{\'{\i}}a}, {Mathur}, {Salabert},
  {Ballot}, {R{\'e}gulo}, {Metcalfe}, \& {Baglin}}]{garcia10}
{Garc{\'{\i}}a}, R.~A., {Mathur}, S., {Salabert}, D., {et~al.} 2010, Science,
  329, 1032

\bibitem[{{Hauschildt} {et~al.}(1999){Hauschildt}, {Allard}, \&
  {Baron}}]{hauschildt99}
{Hauschildt}, P.~H., {Allard}, F., \& {Baron}, E. 1999, \apj, 512, 377

\bibitem[{{Krzesinski} {et~al.}(2012){Krzesinski}, {Baran}, {Winiarski},
  {Fox-Machado}, {Dr{\'o}{\.z}d{\.z}}, {Siwak}, \&
  {Kozie{\l}-Wierzbowska}}]{krzesinski12}
{Krzesinski}, J., {Baran}, A.~S., {Winiarski}, M., {et~al.} 2012, \actaa, 62,
  201

\bibitem[{{MacDonald} \& {Mullan}(2013)}]{macdonald13}
{MacDonald}, J. \& {Mullan}, D.~J. 2013, \apj, 765, 126

\bibitem[{{Maiolino} {et~al.}(2013){Maiolino}, {Haehnelt}, {Murphy}, {Queloz},
  {Origlia}, {Alcala}, {Alibert}, {Amado}, {Allende Prieto}, {Ammler-von Eiff},
  {Asplund}, {Barstow}, {Becker}, {Bonfils}, {Bouchy}, {Bragaglia}, {Burleigh},
  {Chiavassa}, {Cimatti}, {Cirasuolo}, {Cristiani}, {D'Odorico}, {Dravins},
  {Emsellem}, {Farihi}, {Figueira}, {Fynbo}, {Gansicke}, {Gillon},
  {Gustafsson}, {Hill}, {Israelyan}, {Korn}, {Larsen}, {De Laverny}, {Liske},
  {Lovis}, {Marconi}, {Martins}, {Molaro}, {Nisini}, {Oliva}, {Petitjean},
  {Pettini}, {Recio Blanco}, {Rebolo}, {Reiners}, {Rodriguez-Lopez}, {Ryde},
  {Santos}, {Savaglio}, {Snellen}, {Strassmeier}, {Tanvir}, {Testi}, {Tolstoy},
  {Triaud}, {Vanzi}, {Viel}, \& {Volonteri}}]{maiolino13}
{Maiolino}, R., {Haehnelt}, M., {Murphy}, M.~T., {et~al.} 2013, ArXiv e-prints,
  1310.3163

\bibitem[{{Moya} \& {Garrido}(2008)}]{moya08}
{Moya}, A. \& {Garrido}, R. 2008, \apss, 316, 129

\bibitem[{{Moya} {et~al.}(2004){Moya}, {Garrido}, \& {Dupret}}]{moya04}
{Moya}, A., {Garrido}, R., \& {Dupret}, M.~A. 2004, \aap, 414, 1081

\bibitem[{{Mullan}(2010)}]{mullan10}
{Mullan}, D.~J. 2010, Physics of the Sun: A First Course (Chapman and Hall, CRC
  Press, Boca Raton, FL)

\bibitem[{{Quirrenbach} {et~al.}(2012){Quirrenbach}, {Amado}, {Seifert},
  {S{\'a}nchez Carrasco}, {Mandel}, {Caballero}, {Mundt}, {Ribas}, {Reiners},
  {Abril}, {Aceituno}, {Alonso-Floriano}, {Ammler-von Eiff}, {Anglada-Escude},
  {Antona Jim{\'e}nez}, {Anwand-Heerwart}, {Barrado y Navascu{\'e}s},
  {Becerril}, {Bejar}, {Benitez}, {Cardenas}, {Claret}, {Colome},
  {Cort{\'e}s-Contreras}, {Czesla}, {del Burgo}, {Doellinger}, {Dorda},
  {Dreizler}, {Feiz}, {Fernandez}, {Galadi}, {Garrido}, {Gonz{\'a}lez
  Hern{\'a}ndez}, {Guardia}, {Guenther}, {de Guindos}, {Guti{\'e}rrez-Soto},
  {Hagen}, {Hatzes}, {Hauschildt}, {Helmling}, {Henning}, {Herrero}, {Huber},
  {Huber}, {Jeffers}, {Joergens}, {de Juan}, {Kehr}, {Klutsch}, {K{\"u}rster},
  {Lalitha}, {Laun}, {Lemke}, {Lenzen}, {Lizon}, {L{\'o}pez del Fresno},
  {L{\'o}pez-Morales}, {L{\'o}pez-Santiago}, {Mall}, {Martin},
  {Mart{\'{\i}}n-Ruiz}, {Mirabet}, {Montes}, {Morales}, {Morales Mu{\~n}oz},
  {Moya}, {Naranjo}, {Oreiro}, {P{\'e}rez Medialdea}, {Pluto}, {Rabaza},
  {Ramon}, {Rebolo}, {Reffert}, {Rhode}, {Rix}, {Rodler}, {Rodr{\'{\i}}guez},
  {Rodr{\'{\i}}guez L{\'o}pez}, {Rodr{\'{\i}}guez P{\'e}rez}, {Rodriguez
  Trinidad}, {Rohloff}, {S{\'a}nchez-Blanco}, {Sanz-Forcada}, {Sch{\"a}fer},
  {Schiller}, {Schmidt}, {Schmitt}, {Solano}, {Stahl}, {Storz}, {St{\"u}rmer},
  {Suarez}, {Thiele}, {Ulbrich}, {Vidal-Dasilva}, {Wagner}, {Winkler}, {Xu},
  {Zapatero Osorio}, \& {Zechmeister}}]{quirrenbach12}
{Quirrenbach}, A., {Amado}, P.~J., {Seifert}, W., {et~al.} 2012, in Society of
  Photo-Optical Instrumentation Engineers (SPIE) Conference Series, Vol. 8446

\bibitem[{{Rodr{\'{\i}}guez-L{\'o}pez}
  {et~al.}(2012){Rodr{\'{\i}}guez-L{\'o}pez}, {MacDonald}, \& {Moya}}]{crl12}
{Rodr{\'{\i}}guez-L{\'o}pez}, C., {MacDonald}, J., \& {Moya}, A. 2012, \mnras,
  419, L44

\bibitem[{{Rogers} \& {Nayfonov}(2002)}]{rogers02}
{Rogers}, F.~J. \& {Nayfonov}, A. 2002, \apj, 576, 1064

\bibitem[{{Stellingwerf}(1978)}]{stellingwerf78}
{Stellingwerf}, R.~F. 1978, \aj, 83, 1184

\bibitem[{{Unno} {et~al.}(1989){Unno}, {Osaki}, {Ando}, {Saio}, \&
  {Shibahashi}}]{unno89}
{Unno}, W., {Osaki}, Y., {Ando}, H., {Saio}, H., \& {Shibahashi}, H. 1989,
  {Nonradial oscillations of stars}

\end{thebibliography}


\appendix
\section[]{Tables with the summary of the pulsation analysis depending on the mass and grid}

\begin{table*}
 \centering
  \caption{Summary of periods of oscillation, e-folding time, ages, excitation mechanism and excited modes for completely convective models, masses 0.10~M$_\odot$ to 0.25~M$_\odot$, for each of the calculated grids. See the text for more details.}
  \begin{tabular}{@{}llllllll@{}}
  \hline
            &            & \multicolumn{4}{c}{0.10~M$_\odot$}\\
   Grid & $\alpha$=0.5 & $\alpha$=1.0 & $\alpha$=1.5 & $\alpha$=2.0  & $\alpha$=1.0 ATM & $\alpha$=1.0 ATM & $\alpha$=1.0  ATM \\
        & [Fe/H]=0.0 & [Fe/H]=0.0 & [Fe/H]=0.0 & [Fe/H]=0.0 & [Fe/H]=0.0 & [Fe/H]= -0.5 & [Fe/H]= -1.0  \\
 \hline
 P (min) &   & 5.4~h - 4.3~h & 5.6~h - 4.1~h & 5.4~h - 4.2~h &  &  & 4.5~h - 4.4~h  \\
 $\tau_{\rmn{fold}}$ (Myr) &   & 1.5 - 3.3 & 1.9 - 4.1 & 1.5 - 2.9 &  &  & 1.0 - 1.2  \\
 Age (Myr) &   & 0.8 - 2.4 & 0.6 - 2.2 & 0.6 - 2.0 &  &  & 1.0 - 1.2 \\
 Mech. &    & $\epsilon$ (D) & $\epsilon$ (D) & $\epsilon$ (D) &  &  & $\epsilon$ (D)  \\
 Exc. modes &   & $\ell$0k0 & $\ell$0k0 & $\ell$0k0 &  &  & $\ell$0k0  \\
\hline
\hline
            &            & \multicolumn{4}{c}{0.15~M$_\odot$}\\
   Grid & $\alpha$=0.5 & $\alpha$=1.0 & $\alpha$=1.5 & $\alpha$=2.0  & $\alpha$=1.0 ATM & $\alpha$=1.0 ATM & $\alpha$=1.0  ATM \\
        & [Fe/H]=0.0 & [Fe/H]=0.0 & [Fe/H]=0.0 & [Fe/H]=0.0 & [Fe/H]=0.0 & [Fe/H]= -0.5 & [Fe/H]= -1.0  \\
 \hline
 P (min) & 7.4~h - 6.3~h  & 7.3~h - 6.1~h & 7.0~h - 5.9~h & 6.8~h - 5.6~h &  &  &   \\
 $\tau_{\rmn{fold}}$ (Myr) & 1.5 - 1.8  & 1.2 - 1.6 & 0.9 - 1.5 &  0.8 - 2.0 &  &  &   \\
 Age (Myr) & 0.4 - 1.6  & 0.6 - 1.6 & 0.6 - 1.4 & 0.6 - 1.4 &  &  &  \\
 Mech. & $\epsilon$ (D)   & $\epsilon$ (D) & $\epsilon$ (D) & $\epsilon$ (D) &  &  &   \\
 Exc. modes & $\ell$0k0  & $\ell$0k0 & $\ell$0k0 & $\ell$0k0 &  &  &   \\
\hline
\hline
            &            & \multicolumn{4}{c}{0.20~M$_\odot$}\\
   Grid & $\alpha$=0.5 & $\alpha$=1.0 & $\alpha$=1.5 & $\alpha$=2.0  & $\alpha$=1.0 ATM & $\alpha$=1.0 ATM & $\alpha$=1.0  ATM \\
        & [Fe/H]=0.0 & [Fe/H]=0.0 & [Fe/H]=0.0 & [Fe/H]=0.0 & [Fe/H]=0.0 & [Fe/H]= -0.5 & [Fe/H]= -1.0  \\
 \hline
 P (min) &   & 23 - 24 & 7.4~h/ 22  & 6.9~h/ 22 - 23 &  &  &   \\
 $\tau_{\rmn{fold}}$ (Myr) &   & 586 - 340 & 0.9/ 426 - 413 &  1.2/ 486 - 407 &  &  &   \\
 Age (Myr) &   & 200 - 12\,000 & 1.0/ 500 - 12\,000 & 1.0/ 200 - 12\,000 &  &  &  \\
 Mech. &    & $\epsilon$ (He$^3$) & $\epsilon$ (D) / $\epsilon$ (He$^3$) & $\epsilon$ (D) / $\epsilon$ (He$^3$) &  &  &   \\
 Exc. modes &   & $\ell$0k0 & $\ell$0k0 & $\ell$0k0 &  &  &   \\
\hline
\hline
            &            & \multicolumn{4}{c}{0.25~M$_\odot$}\\
   Grid & $\alpha$=0.5 & $\alpha$=1.0 & $\alpha$=1.5 & $\alpha$=2.0  & $\alpha$=1.0 ATM & $\alpha$=1.0 ATM & $\alpha$=1.0  ATM \\
        & [Fe/H]=0.0 & [Fe/H]=0.0 & [Fe/H]=0.0 & [Fe/H]=0.0 & [Fe/H]=0.0 & [Fe/H]= -0.5 & [Fe/H]= -1.0  \\
 \hline
 P (min) & 9.3~h  & 26 - 27 & 7.5~h / 25 - 26 & 25 - 26 &  &  &   \\
 $\tau_{\rmn{fold}}$ (Myr) & 1.1  & 467 - 106 & 1.9/ 414 - 333 &  404 - 327 &  &  &   \\
 Age (Myr) & 1.0  & 200 - 12\,000 & 1.0/ 200 - 12\,000 & 200 - 12\,000 &  &  &  \\
 Mech. & $\epsilon$ (D) & $\epsilon$ (He$^3$) & $\epsilon$ (D) / $\epsilon$ (He$^3$) &  $\epsilon$ (He$^3$) &  &  &   \\
 Exc. modes & $\ell$0k0  & $\ell$0k0 & $\ell$0k0 & $\ell$0k0 &  &  &   \\
\hline
\end{tabular}
\label{tab:0.10-0.25}
\end{table*}

\begin{table*}
 \centering
  \caption{Summary of periods of oscillation, e-folding time, ages, excitation mechanism and excited modes for the completely to partially convective transition models with mass 0.30~M$_\odot$, for each of the calculated grids. F-b accounts for flux blocking. See the text for more details.}
  \begin{tabular}{@{}llllllll@{}}
  \hline
            &            & \multicolumn{4}{c}{0.30~M$_\odot$}\\
   Grid & $\alpha$=0.5 & $\alpha$=1.0 & $\alpha$=1.5 & $\alpha$=2.0  & $\alpha$=1.0 ATM & $\alpha$=1.0 ATM & $\alpha$=1.0  ATM \\
        & [Fe/H]=0.0 & [Fe/H]=0.0 & [Fe/H]=0.0 & [Fe/H]=0.0 & [Fe/H]=0.0 & [Fe/H]= -0.5 & [Fe/H]= -1.0  \\
 \hline
 P (min) & 10.2~h / 29  & 35 / 29 - 31 & 33 / 28 - 29 & 33 / 28 - 29 & 36 & 32 &  \\
 $\tau_{\rmn{fold}}$ (Myr) & 1.1 / 374 - 320  & 705 / 392 - 186 & 609 / 358 - 265 & 577 / 351 - 259 & 625 & 453  &    \\
 Age (Myr) & 1 / 500 - 12\,000  & 100 / 200 - 12\,000 & 100 / 200 - 12\,000 & 100 / 200 - 12\,000  & 100  & 100  &   \\
 Mech. & $\epsilon$ (D) / $\epsilon$ (He$^3$)   & F-b / $\epsilon$ (He$^3$) & F-b / $\epsilon$ (He$^3$) & F-b / $\epsilon$ (He$^3$) & F-b & F-b &   \\
 Exc. modes & $\ell$0k0  & $\ell$0k0 & $\ell$0k0 & $\ell$0k0 & $\ell$0k0 & $\ell$0k0 &   \\
\hline
\hline
\end{tabular}
\label{tab:0.30}
\end{table*}

\begin{table*}
\setlength{\tabcolsep}{4pt}
 \centering
  \caption{Summary of periods of oscillation, e-folding time, ages, excitation mechanism and excited modes for the partially convective models, masses 0.35~M$_\odot$ to 0.60~M$_\odot$, for each of the calculated grids. F-b accounts for flux blocking. See the text for more details.}
  \begin{tabular}{@{}llllllll@{}}
  \hline
             &            & \multicolumn{4}{c}{0.35~M$_\odot$}\\
   Grid & $\alpha$=0.5 & $\alpha$=1.0 & $\alpha$=1.5 & $\alpha$=2.0  & $\alpha$=1.0 ATM & $\alpha$=1.0 ATM & $\alpha$=1.0  ATM \\
        & [Fe/H]=0.0 & [Fe/H]=0.0 & [Fe/H]=0.0 & [Fe/H]=0.0 & [Fe/H]=0.0 & [Fe/H]= -0.5 & [Fe/H]= -1.0  \\
 \hline
 P (min) & 11.1~h / 28 - 39  & 27 - 42 & 31 - 35 & 31 - 34 & 32 - 36  & 26 - 35 & 36  \\

 $\tau_{\rmn{fold}}$ (Myr) & 1.0 / 666 - 196  & 558 - 163   & 496 - 224   & 468 - 214  & 318 - 175  &  261 - 159 & 26 \\

 Age (Myr) & 1 / 100 - 12\,000   & 100 - 12\,000   & 100 - 12\,000   & 100 - 12\,000   & 200 - 12\,000    & 100 - 12\,000  & 100  \\

 Mech. & $\epsilon$ (D) / F-b    & F-b   & F-b   & F-b   & F-b  & F-b   & F-b  \\

 Exc. modes & $\ell$0k0 / $\ell$0,2,3k0  & $\ell$0,2,3k0 & $\ell$0k0 & $\ell$0k0 & $\ell$0,2k0  & $\ell$0,2,3k0 &   $\ell$2k0 \\
\hline
\hline
            &            & \multicolumn{4}{c}{0.40~M$_\odot$}\\
   Grid & $\alpha$=0.5 & $\alpha$=1.0 & $\alpha$=1.5 & $\alpha$=2.0  & $\alpha$=1.0 ATM & $\alpha$=1.0 ATM & $\alpha$=1.0  ATM \\
        & [Fe/H]=0.0 & [Fe/H]=0.0 & [Fe/H]=0.0 & [Fe/H]=0.0 & [Fe/H]=0.0 & [Fe/H]= -0.5 & [Fe/H]= -1.0  \\
 \hline
 P (min) &  11.3~h / 31 - 59  &  29 - 53 / 1.7~h & 29 - 49 & 29 - 48 &  34 - 37; 1.7~h - 2.5~h & 34 - 55 & 30 - 33 \\
 $\tau_{\rmn{fold}}$ (Myr) & 1.0 / 684 - 186  & 643 - 147 & 620 - 160 & 604 - 155 & 256 - 150 & 355 - 153  & 253 - 129   \\
 Age (Myr) & 1.0 / 50 - 12\,000  & 50 - 12\,000 & 50 - 12\,000 & 50 - 12\,000  &  200 - 12\,000  &  50 - 12\,000  &  100 - 12\,000  \\
 Mech. & $\epsilon$ (D) / F-b   & F-b / $\epsilon$ (He$^3$) & F-b  & F-b  & F-b &  F-b &  F-b  \\
 Exc. modes & $\ell$0k0 / $\ell$0,2,3k0  & $\ell$0,2,3k0 / $\ell$2g1 & $\ell$0,2,3k0 & $\ell$0,2,3k0 & $\ell$0k0; $\ell$1,2g1 & $\ell$0,2k0  & $\ell$0,3k0  \\
\hline
\hline
            &            & \multicolumn{4}{c}{0.45~M$_\odot$}\\
   Grid & $\alpha$=0.5 & $\alpha$=1.0 & $\alpha$=1.5 & $\alpha$=2.0  & $\alpha$=1.0 ATM & $\alpha$=1.0 ATM & $\alpha$=1.0  ATM \\
        & [Fe/H]=0.0 & [Fe/H]=0.0 & [Fe/H]=0.0 & [Fe/H]=0.0 & [Fe/H]=0.0 & [Fe/H]= -0.5 & [Fe/H]= -1.0  \\
 \hline
 P (min) &  33 - 61  & 32 - 54; 1.4~h & 31 - 58 & 31 - 56 & 26 - 45; 1.3~h - 2.0~h  & 35 - 57 &   31 - 45, 1.1~h - 2.2~h  \\
 $\tau_{\rmn{fold}}$ (Myr) & 542 - 161  & 487 - 128 & 465 - 136 & 452 - 133 & 408 - 144 & 543 - 134  & 598 - 111   \\
 Age (Myr) & 50 - 12\,000  & 50 - 12\,000 & 50 - 12\,000 & 50 - 12\,000  &  100 - 12\,000  &  50 - 12\,000  &  20 - 12\,000  \\
 Mech. &  F-b   & F-b  & F-b  & F-b  & F-b / $\epsilon$ (He$^3$) &  F-b &  F-b  \\
 Exc. modes & $\ell$0,2,3k0  & $\ell$02,3k0; $\ell$2g1 & $\ell$0,2,3k0 & $\ell$0,2,3k0  & $\ell$0,2,3k0;  & $\ell$0,2k0  & $\ell$0,2,3k0  \\

 &   &  &  &   &  $\ell$1p1, $\ell$2g1/  &   &   \\

 &   &  &  &   &  $\ell$1g1, $\ell$2g2 &     &    \\

\hline
\hline
            &            & \multicolumn{4}{c}{0.50~M$_\odot$}\\
   Grid & $\alpha$=0.5 & $\alpha$=1.0 & $\alpha$=1.5 & $\alpha$=2.0  & $\alpha$=1.0 ATM & $\alpha$=1.0 ATM & $\alpha$=1.0  ATM \\
        & [Fe/H]=0.0 & [Fe/H]=0.0 & [Fe/H]=0.0 & [Fe/H]=0.0 & [Fe/H]=0.0 & [Fe/H]= -0.5 & [Fe/H]= -1.0  \\
 \hline
 P (min) &  38min - 1.2~h &  35 - 62 / 1.7~h - 1.9~h & 34min - 1.3~h  & 34 - 58 & 1.7~h - 1.9~h   &  &  \\
 $\tau_{\rmn{fold}}$ (Myr) & 405 - 183  & 348 - 104 & 330 - 102 & 320 - 110 & 5975 - 480 &   &    \\
 Age (Myr) & 50 - 12\,000  & 50 - 12\,000 & 50 - 12\,000 & 50 - 12\,000  &  3\,000 - 12\,000  &   &    \\
 Mech. &  F-b   & F-b / $\epsilon$ (He$^3$) & F-b  & F-b & F-b,  $\epsilon$ (He$^3$) &   &    \\
 Exc. modes & $\ell$0,2,3k0  & $\ell$0,2,3k0; $\ell$2g2 / & $\ell$0,2,3k0; $\ell$2g1 & $\ell$0,2,3k0  &  $\ell$1g1, $\ell$2g2 &   &   \\

 &   &  $\ell$1g1 &  &   &  &   &   \\
\hline
\hline
            &            & \multicolumn{4}{c}{0.55~M$_\odot$}\\
   Grid & $\alpha$=0.5 & $\alpha$=1.0 & $\alpha$=1.5 & $\alpha$=2.0  & $\alpha$=1.0 ATM & $\alpha$=1.0 ATM & $\alpha$=1.0  ATM \\
        & [Fe/H]=0.0 & [Fe/H]=0.0 & [Fe/H]=0.0 & [Fe/H]=0.0 & [Fe/H]=0.0 & [Fe/H]= -0.5 & [Fe/H]= -1.0  \\
 \hline
 P (min) &  41~min - 1.3~h  &  32 - 57; 1.5~h - 1.9~h & 21 - 42 & 21 - 41 & 1.5~h - 1.8~h   &  &  \\
 $\tau_{\rmn{fold}}$ (Myr) & 279 - 81  & 230 - 73 & 92 - 85 & 92 - 85 & 2796 - 236 &   &    \\
 Age (Myr) & 50 - 12\,000  & 50 - 12\,000 & 100 - 12\,000 & 100 - 12\,000  &  3\,000 - 12\,000  &   &    \\
 Mech. &  F-b   & F-b / $\epsilon$ (He$^3$) & F-b   & F-b & F-b / $\epsilon$ (He$^3$) &   &    \\
 Exc. modes & $\ell$02,3k0  & $\ell$0,1,3k0; $\ell$2g1-g2/   & $\ell$0,1k0, $\ell$2,3p1 & $\ell$0,2k0, $\ell$2,3p1  &  $\ell$2g2 / $\ell$1g1 &   &   \\
            &             &  $\ell$1g1 &                  &                   &             &   &   \\

\hline
\hline
            &            & \multicolumn{4}{c}{0.60~M$_\odot$}\\
   Grid & $\alpha$=0.5 & $\alpha$=1.0 & $\alpha$=1.5 & $\alpha$=2.0  & $\alpha$=1.0 ATM & $\alpha$=1.0 ATM & $\alpha$=1.0  ATM \\
        & [Fe/H]=0.0 & [Fe/H]=0.0 & [Fe/H]=0.0 & [Fe/H]=0.0 & [Fe/H]=0.0 & [Fe/H]= -0.5 & [Fe/H]= -1.0  \\
 \hline
 P (min) &  26~min - 1.9~h     &  36 - 46; 1.1~h - 2.6~h & 19 - 45 & 42 - 43; 1.8~h &  24 - 29; 1.4~h - 3.3~h  & 1.8~h &  \\
 $\tau_{\rmn{fold}}$ (Myr) & 173 - 42  & 86 - 44 & 52 - 50 & 57 - 61 & 1124 - 158 &  2134 &    \\
 Age (Myr) & 20 - 12\,000  & 50 - 12\,000 & 50 - 12\,000 & 100 - 12\,000  &  2\,000 - 12\,000  &  2\,000  &    \\
 Mech. &  F-b   & F-b / $\epsilon$ (He$^3$) & F-b   & F-b  & F-b / $\epsilon$ (He$^3$) & F-b  &    \\
 Exc. modes & $\ell$0-3k0, $\ell$2,3p1  & $\ell$0,1k0, $\ell$2p1;   & $\ell$0,1k0, $\ell$3p1,2  & $\ell$0k0; $\ell$1g1  &  $\ell$2,3p1; $\ell$1g3, $\ell$2g2/ & $\ell$1g1  &   \\
           &                        &  $\ell$1g3, $\ell$2g2/   &  &   &  $\ell$1g1-g2  &   &   \\

          &                        &  $\ell$1,2g1   &  &   &   &   &   \\
\hline
\hline
\end{tabular}
\label{tab:0.35-0.60}
\end{table*}

\bsp

\label{lastpage}

\end{document}